\documentclass[
    ,final            
  ]
  {aipproc}

\layoutstyle{8x11single}

\def\lsim{\lower 2pt \hbox{$\, \buildrel {\scriptstyle <}\over
         {\scriptstyle \sim}\,$}}
\def\gsim{\lower 2pt \hbox{$\, \buildrel {\scriptstyle >}\over
         {\scriptstyle \sim}\,$}}

\begin{document}

\title{Pulsar Physics and GLAST}


\author{Alice K. Harding}{
  address={Astrophysics Science Division, NASA Goddard Space Flight Center, Greenbelt, MD 20771}
}

\begin{abstract}
Rotation-powered pulsars are excellent laboratories for study of particle acceleration as well 
as fundamental physics of strong gravity, strong magnetic fields, high densities and relativity.  
I will review some outstanding questions in pulsar physics and the prospects for finding answers 
with GLAST LAT observations.  LAT observations should significantly increase the number of detected 
radio-loud and radio-quiet gamma-ray pulsars, including millisecond pulsars, giving much better 
statistics for elucidating population characteristics, will measure the high-energy spectrum and 
the shape of spectral cutoffs and determine pulse profiles for a variety of pulsars of different 
age. Further, measurement of phase-resolved spectra and energy dependent pulse profiles of the 
brighter pulsars should allow detailed tests of magnetospheric particle acceleration and radiation 
mechanisms, by comparing data with theoretical models that have been developed. 
\end{abstract}

\maketitle


\section{INTRODUCTION}

The EGRET telescope on the Compton Gamma-Ray Observatory provided a major increase in our understanding of
rotation-powered pulsars\cite{Thomp04}.  The years since have witnessed a number of further advances that  
include the completion of the Parkes Multibeam survey\cite{Man01}, doubling the 
number of known radio pulsars, observations of unprecedented resolution with the Chandra X-ray telescope,
and discovery of new pulsar wind nebulae (PWN) at TeV energies with HESS.  Deep radio observations of
newly discovered X-ray PWNs have resulted in discovery of young pulsars with very low radio 
luminosities\cite{Camilo04}.
As a result, there are many more young, and potential $\gamma$-ray, pulsars known today than at the
time of EGRET.
In addition, there have been fundamental developments in theoretical models of pulsed high-energy 
emission\cite{Harding05} as well as major progress in defining the global structure of the pulsar 
magnetosphere\cite{Spit06}.  
The stage is now set for the Gamma-Ray Large Area Space Telescope (GLAST), a $\gamma$-ray 
telescope with thirty times the sensitivity of EGRET, to detect possibly more than one hundred new $\gamma$-ray
pulsars and to measure a variety of profiles and spectral characteristics.  It is hoped that some of the 
these fundamental questions of pulsar physics will finally have answers.
  
\section{How and where are particles accelerated in the pulsar magnetosphere?}

The location of the pulsed emission site(s) has been very elusive primarily because we observe only the
pulse profile, the small part of the emission beam that sweeps through our line-of-sight.  Constructing
the full geometry of the emission from only the pulse profile is both non-trivial and non-unique.  
But observing at high-energies has a distinct advantage that the emission patterns map the acceleration sites, 
since the accelerated particles radiate most of their energy very quickly.   Thus, there is hope that with
GLAST LAT observations
we can deduce the geometry of acceleration by studying both details of individual pulse profiles and patterns
of emission in a larger population of $\gamma$-ray pulsars, aided by model predictions. 

The theoretical sites of particle acceleration are determined by the distribution of charge in
the magnetosphere and the geometry of pair creation.  The electric field $E_{||}$ given by 
$\nabla \cdot E_{||}  = 4\pi (\rho  - \rho _{GJ} )$ will
accelerate particles parallel to the magnetic field where the local space charge density $\rho$ departs from the
Goldreich-Julian change density $\rho _{GJ}$.  The high-energy radiation of the accelerated particles can
create electron-positron pairs that can screen the $E_{||}$ at pair formation fronts (PFFs), 
thus limiting the acceleration.  The problem is that we do not know the boundary conditions, which
depend both on the surface physics of the neutron star and on the global current flow pattern.  Global
magnetospheric models\cite{Spit06} can derive the current flow but assume ideal MHD ($\rho  = \rho _{GJ}$ 
everywhere).  However, the pair production that presumably supplies enough charge (at least in the magnetospheres 
of young pulsars) to provide the ideal MHD conditions requires a breakdown of this very condition at acceleration
sites.  Thus, in the absence of magnetospheric models which include pair physics, 
the various acceleration models must depend on assumed boundary conditions.  
Polar cap models\cite{DH96} assume that $E_{||}$ 
develops along open field lines above the neutron star surface, forming a vacuum gap\cite{RS75} 
or a space-charge limited flow (SCLF) accelerator\cite{AS79,MT92}.  
Pairs are produced near the polar cap by interaction of high-energy photons with the strong magnetic field. 
In the polar cap SCLF models, acceleration is slower near the last open field line which is 
assumed to be a boundary where
$E_{||} = 0$.  A narrow slot gap forms\cite{Arons83} where particles cannot produce pairs, 
the $E_{||}$ is unscreened 
and acceleration continues to high altitude\cite{MH04}.  Outer gap accelerator models 
\cite{CHR86,Romani96} focus on regions in the outer magnetosphere that cannot fill with charge,
since they lie along open field lines crossing the null surface, $\vec \Omega \cdot \vec B = 0$, where $\rho _{GJ}$ 
reverses sign.  Charges pulled from the polar cap cannot populate the region
between the null surface and the light cylinder, and a vacuum gap forms. Since the magnetic field is relatively weak 
in the outer gap, pairs are produced by interaction of high-energy photons with X-rays from the hot neutron star
surface.  Both slot gaps and outer gaps can only exist in younger and millisecond pulsars that have sufficient pair 
multiplicity\cite{HMZ02,Zhang04}.

\begin{figure} 
  \includegraphics[width=14.0cm]{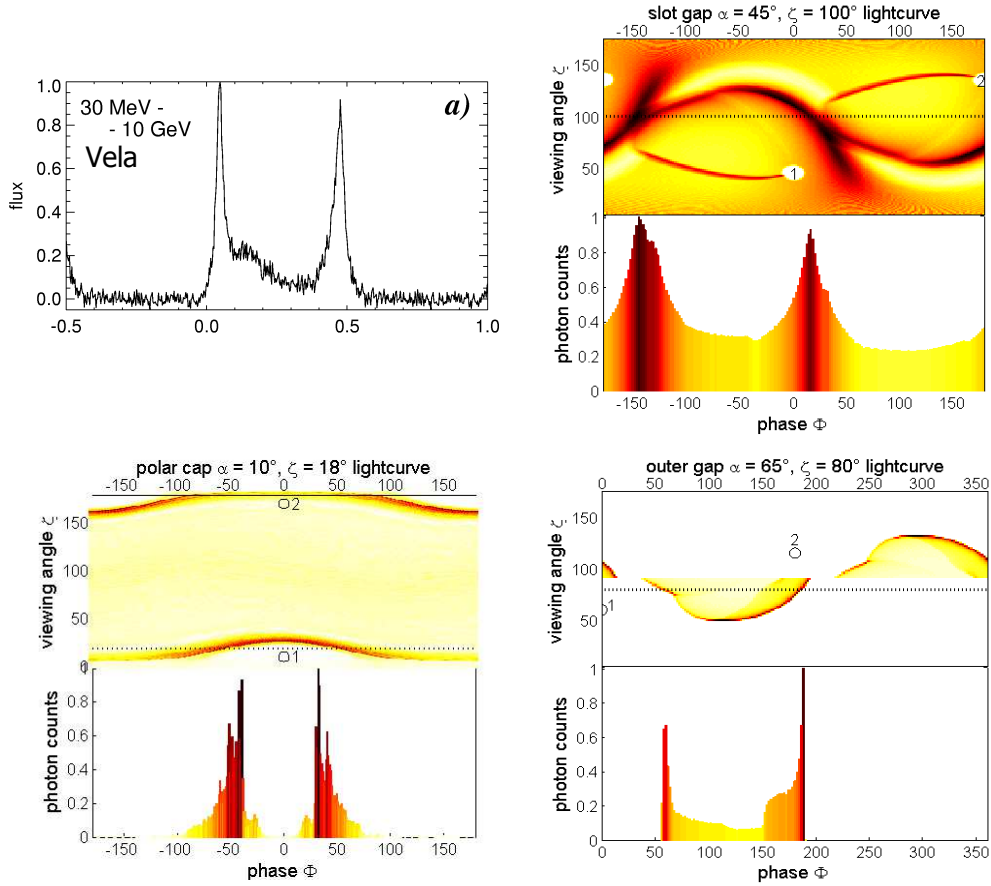}
  \caption{\small Pulse profile of the Vela pulsar, as observed by EGRET, and computed 
phase plots of emission in the ($\zeta_{\rm obs}$, $\phi$) plane and pulse profiles 
produced by a horizontal cut through the phase plot in polar cap, outer gap
and slot gap models\cite{GH06}. $\alpha$ is the pulsar inclination, $\phi$ is the rotational phase and 
$\zeta_{\rm obs}$ is the viewing angle measured from the rotation axis.}
\end{figure}

Figure 1 shows predicted sky emission patterns (phase plots) and pulse profiles for polar cap (PC), slot gap (SG)
and outer gap (OG) models.  Model profiles are generated by cutting through the phase plots at constant 
observer angle. All three models can produce widely spaced, double-peaked profiles similar to that of the Vela pulsar.  
The PC model can generate such profiles only for small observer and inclination angles ($\lsim 10^{\circ}$). 
SG and OG models generate caustics, due to relativistic phase shifts of high-altitude emission, that may be seen
at larger viewing angles(see \cite{Harding05,GH06} for more details).  In the SG model, 
an observer views caustic emission from both poles, whereas in OG models, an observer can view emission 
from only one pole.  Although the model
profiles look similar, there are several distinguishing details that more sensitive measurement may verify.
PC and SG models predict off-pulse emission throughout the entire pulse phase, while OG models do not.  
Recent OG models\cite{Takata07} include emission from both poles, but this adds only a trailing 
edge to the peaks.  The SG model predicts a possible enhancement on the trailing edge of the first peak, 
caused by overlapping field lines near the light cylinder (an effect which forms the first peak in the OG model).  

\vskip 0.2cm
\begin{figure} 
 \includegraphics[width=9cm]{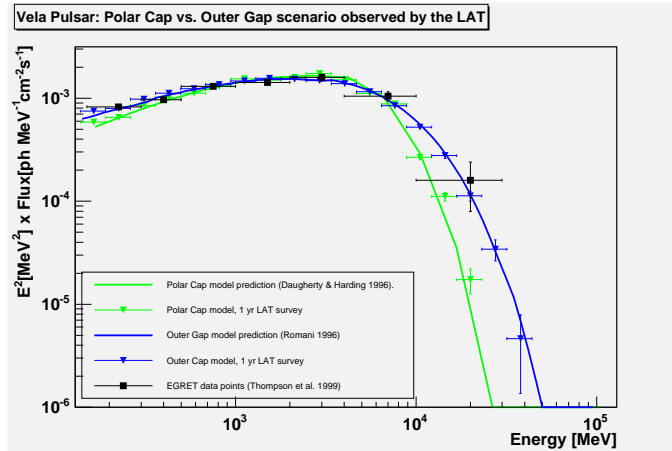}
  \caption{Simulated spectra of the Vela pulsar observed in 1 year by LAT assuming polar cap and outer gap
models\cite{RH07}, compared with observed EGRET spectrum.}
\end{figure}

PC models predict that the pulsed spectrum from polar cap cascades will have a sharp 
(super-exponential) high-energy cutoff due to attenuation by magnetic pair production and photon 
splitting\cite{BH01}.  The energy of these cutoffs is a function of the local 
magnetic field strength, pulsar period and the emission altitude and lies between 1 and 20 GeV.
EGRET detected indications of spectral cutoffs, but did not have the high-energy sensitivity to distinguish
a super-exponential from the simple exponential cutoffs predicted by high-altitude SG and OG models. 
GLAST should be able to measure the shape of the spectral cutoffs well enough to make this distinction and
therefore place strong limits on the altitude of emission.  Simulations of the Vela pulsar spectrum (Figure 2) 
show that in 1 - 2 months of observation, the GLAST LAT will be able to measure the cutoff well enough to
distinguish PC from OG model shapes\cite{RH07}.  If sharp cutoffs are measured in a number 
of pulsar spectra, then one can also look for the predicted correlation between cutoff energy and surface field 
strength.

\section{What are the high-energy radiation mechanisms?}

All models predict that the high-energy spectra are composed of multiple radiation components, although not all
necessarily contribute in the LAT energy range.  
Curvature radiation (CR) from primary particles, synchrotron radiation (SR) 
from pairs and Inverse Compton scattering (ICS) components generally are present for all models.  In addition,
there is a thermal X-ray component that may be hidden for Crab-like pulsars.  
PC pair cascade spectra\cite{DH96} comprise a CR component at high energies that is often 
cutoff by pair production before the pair SR component dominates, although in pair-starved pulsars the 
SR component may be much weaker. 
An ICS component from pairs scattering thermal X-rays from the neutron star surface may appear at 
X-ray energies\cite{ZH00}.  In OG models, the pair SR component is dominant for Crab-like pulsars
and a transition between CR and SR components may be visible.  The predicted ICS component in OG models comes
from primary particles scattering soft photons from various sources and appears above 100 GeV.  Although air Cherenkov telescopes have searched for this component for many years, it was not been 
detected\cite{Aharon07} and it is therefore very unlikely that GLAST will see it.  
The radiation from the high-altitude 
SG consists of a CR component from primary electrons at the highest energies, and a SR component from both 
primaries and pairs from lower altitude cascades that resonantly absorb radio emission at higher 
altitudes\cite{Harding07}.   

GLAST LAT observations should be able to distinguish multiple spectral components in two ways.  Better 
sensitivity and energy resolution will enable detection of transitions in the phase-averaged spectra.  In fact,
the Crab pulsar spectrum seems to have a transition between multiple components at an energy 
around 100 MeV\cite{Kuiper01}
well within the LAT range.  Furthermore, the LAT will obtain phase-resolved spectra having error bars five times smaller than EGRET.  These much improved statistics will enable measurement of spectral components, cutoff energy 
and shape as a function of phase to make more detailed comparison with model predictions.  For example, PC models
predict that the high-energy cutoffs are sharpest at the pulse peaks\cite{DH96} and the cutoff energy is lower in the 
first peak\cite{DR02}.  

\section{Are processes the same for all pulsars?}

It is quite possible that accelerator type, spectral formation, radiation processes and emission altitude depend
on pulsar parameters such as period, age or surface field strength.  To explore this, one can look for patterns
in both multiwavelength profiles and in broadband spectra.  
The EGRET pulsars showed an increase in spectral hardness with age\cite{Thomp94}, causing the
power peak of the broadband spectrum to move from hard X-rays (in the case of the Crab) to high-energy 
$\gamma$-rays for the middle-aged pulsars.  This indicates that the mechanism(s) that contribute to the
non-thermal 
X-ray emission are less dominant in older pulsars.  Most models predict this behavior, since the pair cascade
multiplicity decreases with age, and it is radiation from pairs that contributes most at X-ray energies.
It will be interesting to see if this trend persists in the larger population of pulsars that LAT will detect.

The behavior of profiles across energy bands displays fewer clear patterns.  For most EGRET pulsars, the 
$\gamma$-ray peaks do not resemble those at other wavelengths and are not phase-aligned.  Only the Crab
shows phase-alignment of the peaks at all wavelengths from radio to $\gamma$ rays.  Although a trend is hard to
establish among the EGRET pulsars themselves, if one examines the profiles of X-ray pulsars one sees that
those having short periods (including several millisecond pulsars) show more phase alignment.  
This behavior is not simply related to the patterns in
the broadband non-thermal spectra since it also involves the radio profile alignment, and is more probably
related to emission geometry.  For example, the smaller magnetospheres of pulsars with shorter periods may
put more of the emission at different wavelengths at high altitudes (relative to the light cylinder) so that
caustic effects will align all the emission in phase\cite{Harding05}.  

\section{Are there $\gamma$-ray millisecond pulsars?}

A number of studies have predicted that millisecond pulsars should have a high-energy CR radiation component
originating from polar cap acceleration and extending above 10 GeV\cite{Bulik00,Luo00,HVM05}.  Because of their very low surface magnetic fields, the pair production attenuation 
cutoffs are at much higher energies than in normal pulsars and typically occur above 100 GeV.  The sensitivity
of the GLAST LAT to these high energies will make millisecond pulsars ideal sources, but how many will be detectable will depend on whether our viewing angle sees much of the low altitude CR.  The presence of pulsed X-ray emission,
as many millisecond pulsars show, may not necessarily be a guide since the X-rays may originate from higher 
altitude\cite{HVM05}.  
EGRET marginally detected the millisecond radio and X-ray pulsar J0218+4232, but the spectrum
is very soft.  GLAST LAT observations will certainly confirm this detection and give a clearer picture of the 
$\gamma$-ray spectrum.  CR emission from particles accelerating in the outer gaps of ms pulsars cannot extend above 
several GeV in present models (Hirotani, priv. comm.) 
so that detected emission above 10 GeV from these sources must originate
from near the polar cap.  However, the cascades from outer gap-accelerated particles moving downward toward the 
polar cap may radiate CR above 10 GeV\cite{ZC03}.  But in this case, one would expect that the
$\gamma$-ray pulses will be out of phase with the radio pulses.

\section{What is the ratio of radio-loud to radio-quiet $\gamma$-ray pulsars?}

The number of radio-loud and radio-quiet pulsars that GLAST detects after a 1 to 2 year survey will give  
us strong constraints on the $\gamma$-ray emission geometry and type of accelerators that are operating.
PC, SG and OG models give very different predictions for the ratio of the radio-loud to radio-quiet populations,
so estimating this ratio will be a sensitive diagnostic.  In the population synthesis studies, the term
radio-quiet refers to $\gamma$-ray sources that are not detectable by current radio surveys, either because
the radio beam is just not visible at our viewing angle or because the radio emission is absent or very weak.
PC models predict the highest ratio of radio-loud to radio-quiet pulsars, since the $\gamma$-ray and radio beams
are co-aligned with the magnetic axis.  High-altitude emission models predict much lower ratios and many more
radio-quiet pulsars, since the $\gamma$-ray beams are most often counter-aligned with the radio beams.  
Current predictions are shown in Table 1.  The results of these studies indicate that if the LAT finds 
many of the EGRET unidentified source near the Galactic plane are radio-loud pulsars, then the emission
must come from a low-altitude accelerator.  Many of the simulated radio-quiet millisecond pulsars are located 
outside the bounds of the surveys, but are bright enough to possibly be detected by follow-up radio observations.

\begin{table}
\begin{tabular}{lcccc}
\hline
& \multicolumn{2}{c}{\bf Normal pulsars} & \multicolumn{2}{c}{\bf Millisecond pulsars} \\
& Radio-loud & Radio-quiet & Radio-loud & Radio-quiet \\
\hline
Low Altitude Slot gap \cite{Gonthier06}\cite{Story06} & 81 & 43 & 16 & 99-131 \\
High Altitude Slot gap \cite{Harding06} & 5 & 49 & &  \\
Outer Gap \cite{Harding06} & 1 & 258 && \\
~~~~~~~~~~\cite{Jiang06} & 78 & 740 && \\
\hline
\end{tabular}
\caption{Predicted GLAST pulsar populations}
\end{table}

\section{Summary}

GLAST LAT observations promise to provide answers to some fundamental questions about pulsar acceleration
and emission, including the location and type of accelerator and the dominant modes of emission.  
In addition to detecting many more radio pulsars in $\gamma$ rays, GLAST will detect previously unknown 
radio-quiet pulsars as well as aid in detecting new millisecond pulsars.




\bibliographystyle{aipproc}   

\begin{thebibliography}{9}

\bibitem{Aharon07}
F. Aharonian et al. \emph{A \& A}, submitted [astro-ph0702336] (2007)

\bibitem{Arons83} 
J. Arons, \emph{ApJ}, 266, 215 (1983)

\bibitem{AS79}
J. Arons \& E. T. Scharlemann, \emph{ApJ}, 231, 854 (1979)

\bibitem{BH01}
M. G. Baring \& A. K. Harding, \emph{ApJ}, 547, 929 (2001)

\bibitem{Bulik00}
T. Bulik, B. Rudak \& J. Dyks, \emph{MNRAS}, 317, 97 (2000) 

\bibitem{Camilo04}
F. Camilo,  in ``Young Neutron Stars and Their Environments", IAU Symp. 218, Ed, F. Camilo \& B. M. Gaensler. San Francisco, CA: Astronomical Society of the Pacific, 2004., p.97

\bibitem{CHR86}
K.~S. Cheng, C. Ho, \& M.~A. Ruderman, \emph{ApJ}, 300, 500, 1986.

\bibitem{DH96}
J. K. Daugherty  \& A. K. Harding, \emph{ApJ}, 458, 278 (1996) 

\bibitem{DR02}
Dyks, J. \& Rudak, B. \emph{A \& A}, 393, 511 (2002)

\bibitem{Gonthier06}
Gonthier, P.L., Story, S.A., Clow, B. D. \& Harding, A.K., \emph{A \& A}, in press [astro-ph0702097] (2007)

\bibitem{GH06}
I. A. Grenier \& A. K. Harding, in Proc. of  Einstein Centenary Conference, Paris 2005, in press (2006)

\bibitem{Harding05}
A. K. Harding,  in Proc. of 22nd Texas Symp. on Rel. Astrophys.,
ed. P.Chen et al., econf C041213 (astro-ph/0503300) (2005)

\bibitem{Harding06}
A. K. Harding, I. A. Grenier \& P. L. Gonthier,\emph{A \& A}, in press [astro-ph0703019] (2007)

\bibitem{HMZ02}
A. K. Harding, A. G. Muslimov \& B. Zhang, \emph{ApJ} {\bf 576} 366 (2002)

\bibitem{HVM05}
A. K. Harding, V. V. Usov, \& A. G. Muslimov, \emph{ApJ}, 622, 531 (2005)

\bibitem{Harding07}
A. K. Harding, J. V. Stern, J. Dyks \& M. Frackowiak, in preparation (2007)

\bibitem{Jiang06}
Z. J. Jiang \& L. Zhang, \emph{ApJ}, {\bf 643}, 1130 (2006)

\bibitem{Kuiper01}
L. Kuiper et al. \emph{A \& A}, 378, 918 (2001)

\bibitem{Luo00}
Q. Luo, S. Shibata \& D. B. Melrose, \emph{MNRAS}, 318, 943 (2000)

\bibitem{Man01}
R. N. Manchester et al. \emph{MNRAS}, 328, 17 (2001)

\bibitem{MH04}
A. G. Muslimov, \& A. K. Harding, \emph{ApJ}, 606, 1143 (2004)

\bibitem{MT92}
A. G. Muslimov \& A. I. Tsygan, \emph{MNRAS}, 255, 61 (1992)

\bibitem{RH07}
M. Razzano \& A. K. Harding, these proceedings (2007)

\bibitem{Romani96}
R.~W. Romani, 1996, \emph{ApJ}, 470, 469.

\bibitem{RS75}
M. A. Ruderman \& P. G. Sutherland, 1975, \emph{ApJ}, 196, 51.

\bibitem{Spit06}
A. Spitkovsky, \emph{ApJ}, 648, L51 (2006)

\bibitem{Story06}
Story, S.A., Gonthier, P.L. \& Harding, A.K. \emph{ApJ}, these proceedings (2007)

\bibitem{Takata07}
Takata, J.; Chang, H.-K.; Cheng, K. S. \emph{ApJ}, 656, 1044 (2007)

\bibitem{Thomp04}
Thompson, D. J. in {\it Cosmic Gamma-Ray Sources}, ed. K. S. Cheng \& G. E. Romero, (Kluwer) p. 149 (2004)

\bibitem{Thomp94}
Thompson, D. J. et al. \emph{ApJ}, 436, 229 (1994)

\bibitem{ZC03}
L. Zhang \& K. S. Cheng, \emph{A \& A}, 398, 639 (2003)

\bibitem{ZH00}
B. Zhang, \& A. K. Harding, {\it ApJ} {\bf 535} L51 (2000)

\bibitem{Zhang04}
L. Zhang, K. S. Cheng, Z. J. Jiang,  P. Leung, \emph{ApJ}, {\bf 604}, 317 (2004)

\end{thebibliography}



\end{document}